\documentclass[preprint,nofootinbib,preprintnumbers,amsmath,amssymb]{revtex4}
\usepackage[dvips]{graphicx}
\usepackage{graphicx}
\usepackage[hidelinks]{hyperref}
\usepackage{amsfonts}
\usepackage{wrapfig}
\usepackage{bm}
\usepackage{amsmath}
\usepackage{amssymb}
\usepackage{color}
\usepackage[all]{xy}
\usepackage{subfigure}
\usepackage{float}  
\usepackage{here}  
\usepackage{latexsym}         
\usepackage{graphicx}
\usepackage{wrapfig}
\providecommand{\abs}[1]{\lvert#1\rvert}
\begin{document}

\title{Extended thermodynamics and critical behavior of  generalized dilatonic Lifshitz black holes }

\author{Carlos E. Romero-Figueroa$^1$, and Hernando Quevedo$^{1,2,3}$} 
\email{carlosed.romero@correo.nucleares.unam.mx;quevedo@nucleares.unam.mx}

\affiliation{$^1$Instituto de Ciencias Nucleares, Universidad Nacional Aut\'onoma de M\'exico, AP 70543, Mexico City, Mexico}
\affiliation{$^2$Dipartimento di Fisica and Icra, Universit\`a di Roma “La Sapienza”, Roma, Italy.}
\affiliation{$^3$Al-Farabi Kazakh National University, Al-Farabi av. 71, 050040 Almaty, Kazakhstan.}

\date{\today}

\begin{abstract}
We study a particular  Einstein-Maxwell-Dilaton black hole configuration with cosmological constant, expressed in terms of the curvature radius,  from the point of view of quasi-homogeneous thermodynamics. In particular, we show that the curvature radius and the coupling constant of the matter fields can be treated as thermodynamic variables in the framework of extended thermodynamics, leading in both cases to a van der Waals-like behavior. We also investigate in detail the stability and critical properties of the 
black holes and obtain results, which are compatible with the mean field approach. 


\end{abstract}
\maketitle
\section{Introduction}
From ordinary thermodynamics, it is known that to fully describe the thermodynamic properties of a system, we need a homogeneous fundamental equation that relates all the extensive quantities of the system \cite{callen1998thermodynamics}, for which there exist two representations, namely, the energy and entropy representations, depending on which potential is used to describe the system.  In the thermodynamics of non-ordinary systems such as black holes, we want to inherit this neat feature of standard thermodynamics. However, in contrast to ordinary laboratory systems,  black holes in general cannot be properly described by homogeneous fundamental equations \cite{quevedo2019quasi}. Nevertheless, this can be fixed, if the fundamental equation belongs to a family of more general kinds of functions, the so-called quasi-homogeneous functions \cite{quevedo2017homogeneity,quevedo2019quasi}. Quasi-homogeneous functions were introduced into the framework of standard thermodynamics to study scaling and universality near the critical point \cite{hankey1972systematic,neff1974generalization}. In fact, the idea of considering quasi-homogeneous thermodynamics,
particularly in black hole physics and in geometric representations of thermodynamics, has been previously analyzed by Belgiorno and Cacciatore in a series of publications \cite{belgiorno2003quasi,belgiorno2003black,belgiorno2011general}. It is interesting that quasi-homogeneity is also present in the fundamental equations of other 
self-gravitating systems, such as non-relativistic rotating fermionic matter and self-gravitating radiation \cite{chavanis2003statistical}. Quasi-homogeneous behavior in thermodynamics has been shown to correctly describe at least black holes of the Kerr-Newman family \cite{belgiorno2003quasi,belgiorno2003black}, the Hertel-Narnhofer-Thirring (HNT) model for self-gravitating fermions \cite{thirring1972z} 
and spherical thermal geons in \cite{power1957thermal}.\vspace{0.1in}

 It is important to mention that mathematically it is possible to convert a quasi-homogeneous function into a homogeneous function of the first degree by making an appropriate change of variables, but in non-ordinary systems such as black holes, it seems that this drastically alters the physical properties of the system. For example, in Kerr-Newman black holes the phase transition structure is modified, which leads to contradictory results in their thermodynamics \cite{quevedo2019quasi,quevedo2023unified}. Furthermore, to consider black holes as quasi-homogeneous systems, an extended thermodynamic space is required. For example,  in the case of asymptotically AdS black hole solutions, the cosmological constant has been interpreted as a new thermodynamic variable with properties consistent with an effective “pressure” \cite{kubizvnak2017black,dolan2012pdv,kastor2009enthalpy}. According to these results, the mass of the black hole no longer has the meaning of internal energy. Instead, it should be interpreted as the gravitational version of chemical enthalpy \cite{kastor2009enthalpy}. Similar results have already been reported in the case of Lovelock gravitational theories \cite{lovelock1971einstein}, where the curvature coupling constants appear explicitly in the Smarr relation \cite{jacobson1993entropy,kastor2010smarr}. A similar situation occurs in the Born-Infeld theory of non-linear electrodynamics \cite{gunasekaran2012extended}, where the coupling parameter that represents the maximum strength of the electromagnetic field adds an extra term to the Smarr relation and whose conjugate potential is interpreted as an {electrical polarization of the vacuum}. These results have been generalized to higher dimensions, and there has been further research on the extended equilibrium space in Born-Infeld theories \cite{zou2014critical,hendi2014extended,hendi2016p,mo2014p,belhaj2015ehrenfest,hendi2015extended}.

In this work, we will study a particular black hole configuration from the point of view of quasi-homogeneous thermodynamics. The black hole spacetime is given by a solution of a generalization of Einstein-Maxwell gravity, which includes a dilatonic field non-minimally coupled to the electromagnetic  field. In addition, the spacetime metric is assumed to be asymptotically Lifshitz, leading to a rich geometric and physical structure  as a result of the anisotropic properties of the background  spacetime.  In Sec. \ref{section2}, we briefly review the  main properties of the black hole solution, emphasizing the role of the coupling constants and the essential thermodynamic properties. In Sec. \ref{sec:quasi}, we explicitly identify the black hole spacetime as a quasi-homogeneous thermodynamic system and determine the parameters that should be identified as thermodynamic variables. 
In Sec. \ref{ko}, we analyze the black hole system in the framework of extended  thermodynamics, considering the coupling constant of the matter fields and the cosmological constant, expressed in terms of the curvature radius, as independent thermodynamic variables. Then, in Sec. \ref{sta}, using the standard approach of classical equilibrium thermodynamics, we present a detailed analysis of the stability properties of the system from a local and a global point of view. This analysis is followed by an investigation of the critical properties of the generalized Lifshitz black holes. In Sec. \ref{sec:crit}, we perform a detailed derivation of the critical exponents and of the Gibbs free energy, from which the phase transition structure of the black holes is determined. 
Finally, in Sec. \ref{sec:con}, we summarize our results.

\section{Generalized Einstein-Maxwell-Dilaton Lifshitz black holes}
\label{section2}

The generalized Einstein-Maxwell-Dilaton (EMD) framework is described by the following $D$-dimensional gravitational action
\begin{equation}
 \label{accion1}
S[g,A,\phi]= \int d^{D}x\sqrt{-g}\left(\frac{R-2\Lambda}{2\kappa}- \frac{1}{2}\partial_\mu\phi\partial^\mu\phi-\frac{1}{4}h(\phi)F_{\mu\nu}F^{\mu\nu}\right).
\end{equation}
 Following  standard notation, $\sqrt{-g}$ is the square root of the determinant of the  metric, $R$ is the scalar curvature, $\Lambda$ is the cosmological constant, and $\kappa\equiv 8 \pi G$ is the $D$-dimensional Einstein's gravitational constant. The matter fields correspond to a real scalar field $\phi$ and a non-massive vector field $A_\mu$. In addition, 
 $h(\phi)$ is a generalized coupling function of the matter fields  (when $h(\phi)=e^{-\alpha\phi}$, for $\alpha$ constant, the standard dilatonic coupling is recovered \cite{taylor2008non}).
 We will study $D$-dimensional static, planar Lifshitz-like black hole solutions, which are described by the following  finite temperature metric
\begin{equation}
\label{metrica1}
    ds^2=-\left(\frac{r}{l}\right)^{2z}f(r)dt^2+\frac{l^2dr^2}{r^2 f(r)}+\left(\frac{r}{l}\right)^{2}\sum^{D-2}_{i=1} dx^2_i,
\end{equation}
 where $l$ is the curvature radius of the spacetime, $z$ is the dynamical critical exponent\footnote{In the limit $z\rightarrow 1$, we recover  the AdS metric.} and the $r$-coordinate is an extra dimension that is identified with energy in the dual holographic theory. The above metric  with
\begin{equation}
 f(r)=1-\frac{m}{r^{D+z-2}}+\frac{b}{r^{2(D+z-3)}},\quad m\equiv{r_h}^{D+z-2}\left(1+br_h^{-2(D+z-3)}\right),\label{mko}
\end{equation}
is a solution to the field equations \cite{ herrera2021scalarization},
where $r_h$ is the position of the event horizon, $m$ is the mass parameter, and $b$ is an integration constant that generalizes the dilatonic coupling. The most general coupling function is
\begin{equation}
 h(\phi)=\frac{-4(D+z-3)\kappa q^2l^{2(D-1)}e^{-2\sqrt{\frac{\kappa(D-2)}{(z-1)}}(\phi-\phi_0)}}{2(z-1)\Lambda l^2-bl^{-2(D+z-3)}(D-2)(D+z-3)(D+z-4)e^{-2\sqrt{\frac{\kappa}{(z-1) (D-2)}}(\phi-\phi_0)}},\label{h}
\end{equation}
where $q$ is the charge density associated with the Maxwell field. Additionally, we have the standard $D$- dimensional AdS cosmological constant,  a logarithmic scalar field, and a purely electrical vector field, given by
\begin{equation}
\phi(r)=\sqrt{\frac{(D-2)(z-1)}{\kappa}{}}\ln\left(\frac{r}{l}\right)+\phi_o,\quad \Lambda=-\frac{(D+z-3)(D+z-2)}{2l^2},\label{cosmo}
\end{equation}
\begin{equation}
    A(r)=\frac{(z-1)}{4 \kappa q l^3}\bigg(\frac{r}{l}\bigg)^{(D+z-2)}-\frac{b(D-2)}{4 \kappa q l^{2(D+z)-5}}\bigg(\frac{l}{r}\bigg)^{(D+z-4)}. \label{max}
\end{equation}

\begin{figure}[H]
\begin{minipage}[t]{0.47\linewidth}
 \centering
\hspace{1cm}
\includegraphics[width=1\linewidth]{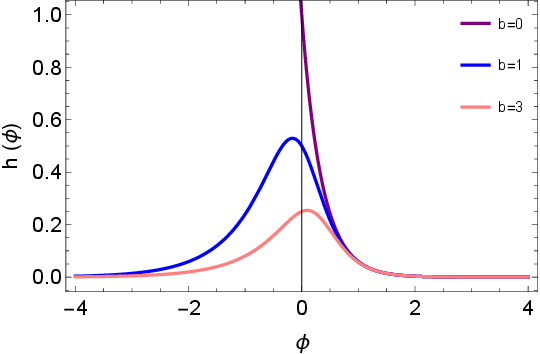}
 (a)\hspace{10cm}
\end{minipage}%
\hfill%
\begin{minipage}[t]{0.49\linewidth}
 \centering
\hspace{1cm}
\includegraphics[width=1\linewidth]{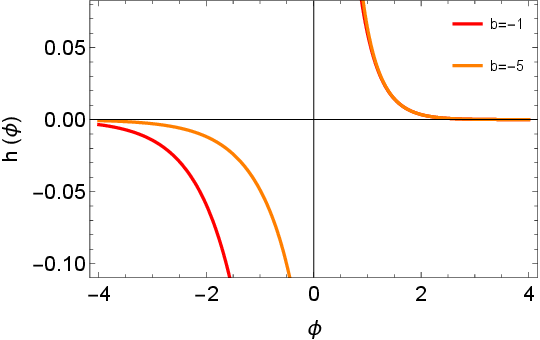}
(b)\hspace{10cm}
\end{minipage}
    \caption{Coupling function $h(\phi)$ for fixed $l=1$, $\kappa=1$, $q=1$ $D=4$, $z=2$, and $\phi_0=0$. (a) $h(\phi)$ for $b\geq 0$. (b) $h(\phi)$ for $b<0$. A similar behavior is observed for other  values of $D$, $l$, and $z$.}\label{h1}
\end{figure}

To understand the role of the coupling constant $b$, we plot in Fig. \ref{h1} the behavior of $h(\phi)$ for different values of $b$. Fig. \ref{h1} shows that for $b>0$ the coupling function behaves like $h(\phi)\propto 1/cosh \phi$. However, for $b<0$ the coupling behaves like $h(\phi)\propto 1/sinh \phi$. Moreover, in both cases, with the increase of $b$ the effect of the Maxwell invariant becomes smaller and smaller. When $\abs b\rightarrow \infty$, gravity takes over electromagnetic interaction and the electromagnetic field can be safely neglected. This means the value of $b$ determines the strength of interaction between the Maxwell field and the scalar field. 

It is observed that the obtained coupling can be understood as a nonlinear deviation\footnote{When the parameter $b\rightarrow 0$, the standard EMD theory is recovered and the entire configuration is reduced exactly to the Taylor black hole \cite{taylor2008non}.} from the standard dilaton coupling, measured by the order of the new parameter $b$. It is well-known that for Lifshitz planar black holes in a standard dilatonic setup, we need at least two vector fields to obtain a charged solution \cite{tarrio2011black,brenna2015mass}. Nevertheless, in the black hole described by the field configuration Eqs.(\ref{mko})-(\ref{max}), the parameter $b$ appears as a free charge in the metric and it seems that the entire configuration behaves like a charged black hole, once we identify $b$ as a normalized free charge square of a second vector field with charge density $\Tilde{q}$, namely
\begin{equation}
    b\equiv \frac{\Tilde{q}^2l^{2z}\mu^{-3\sqrt{\frac{(D-2)}{(z-1)}}}}{2(D-2)(D+z-4)},\label{xs}
\end{equation}
where $\mu\equiv l^{-\sqrt{(D-2)(z-1)}}$ is the amplitude of the scalar field Eq.(\ref{cosmo}). In the above expression, we modified the power of the amplitude of the scalar field $\mu$, for the parameter $b$ has the right units\footnote{Notice that in \cite{brenna2015mass} the matter fields are dimensionless and $[q^2]=L^{2(D-2-z)}$. On the contrary, we are using here  units with  $[A_\nu]=L^{-D/2+1}$, $[\phi]=L^{-D/2+1}$, and $[q^2]=L^{-D}$.} \cite{tarrio2011black}, i.e., $[b]=L^{2(D+z-3)}$. This identification will come into play later when we analyze the extended thermodynamic equilibrium space of the EMD black hole. We can spot from the field configuration Eqs.(\ref{mko})-(\ref{max}) that in the AdS limit, i.e., $z=1$, the dilaton field reduces to its ground value and one branch of the electric field Eq.(\ref{max}) is zero. Additionally, if we make the identification Eq.(\ref{xs}), in the isotropic limit the entire configuration reduces to a planar Reissner-Nordström-AdS (pRN-AdS) black hole solution considered in \cite{chamblin1999charged}.
\\ \\
 The Hawking temperature associated
with the $D$-dimensional configuration  reads
\begin{equation}
    T_H=\frac{1}{4\pi}\frac{(D+z-2)r_h^z-b(D+z-4)r_h^{-(2D+z-6)}}{l^{z+1}},\label{temp}
\end{equation}
 and the Wald entropy results proportional to the event horizon area $A$ and takes the form
\begin{equation}
    S=\frac{A}{4 G}=\frac{2 \pi w}{\kappa}r_h^{(D-2)},\label{entro}
\end{equation}
 where $w\equiv l^{2-D}\int d^{D-2}x$ represents the dimensionless Euclidean volume of the spatial sector. The electric charge can be computed through a Gaussian integral over a spatial hypersurface at asymptotic
infinity 
\begin{equation}
    Q_e=\int_\Sigma h(\phi)\star F= q V_{D-2}\label{carga}.
\end{equation}
Next, in order to compute the mass of the black hole configuration the so-called
generalized ADT quasilocal method was exploited, foremost introduced in \cite{kim2013quasilocal}. For this configuration, we obtain 
\begin{equation}
\mathcal{M}_{quasilocal}=\frac{(D-2)w}{2\kappa l^{z+1}}m,
    \label{masa}
\end{equation}
 Notice that $\mathcal{M}_{quasilocal}$ coincides with the mass calculated in \cite{tarrio2011black} by using a Komar integral with the black hole solution, subtracting the value from the thermal case ($m = 0$ and $q = 0$). Moreover, if we make the identification Eq.(\ref{xs}) in the scalar-free limit $z\rightarrow1$, the quasi-local mass reduces to the ADM mass of a pRN-AdS black hole solution, and additionally for $b=0$ the configuration reduces to a planar Schwarzschild AdS black hole \cite{brenna2015mass}.

\section{Lifshitz black holes as quasi-homogeneous systems}
\label{sec:quasi}

Let $\Phi$ denote a fundamental thermodynamic potential \cite{callen1998thermodynamics} which could be either the
entropy or the internal energy. Let ${E^a}$ ($a = 1, . . . , n$) denote the set of extensive variables that are necessary to describe a thermodynamic system with $n$ degrees of freedom. Then, a system described by the fundamental equation $\Phi(E^a)$ is called ordinary if $\Phi$ is a homogeneous
function
\begin{equation}
    \Phi(\lambda E^a)=\lambda^\beta\Phi(E^a),
\end{equation}
where $\lambda$ is a real constant and $\beta > 0$ is the degree of homogeneity. In general, ordinary systems are characterized by the value $\beta = 1$. If $\Phi$ is a quasi-homogeneous function \cite{quevedo2019quasi}, i.e.,
\begin{equation}
    \Phi(\lambda^{\beta_1} E^1,...,\lambda^{\beta_n} E^n)=\lambda^{\beta_\Phi}  \Phi( E^a), \label{homogen2}
\end{equation}
where $\beta_a=(\beta_1,...,\beta_n)$ are real constants, and $\beta_\Phi$ is the generalized degree of homogeneity, the system
is called non-ordinary. In particular for the Lifshitz black hole described in Sec. \ref{section2} it is trivial to obtain the fundamental equation in the energy representation, and read
\begin{equation}
     M=\frac{(D-2)w}{2\kappa l^{z+1}}\left[ \Bigg(\frac{\kappa S}{2 \pi w}\Bigg)^{(D+z-2)/(D-2)}+b\Bigg(\frac{\kappa S}{2 \pi w}\Bigg)^{-(D+z-4)/(D-2)}\right].
    \label{fundame11}
\end{equation}
Notice that Eq.(\ref{fundame11}) is not a homogeneous function. However, if we interpret the coupling parameter $b$ as thermodynamic variable, the fundamental equation is a quasi-homogeneous function $M(\lambda^{\beta_S}S,\lambda^{\beta_b}b)=\lambda^{\beta_M} M(S,b) $ of degree $\beta_M=(D+z-2)/(D-2)\beta_S$, if the condition
\begin{equation}
    \beta_b=\frac{2(D+z-3)}{(D-2)}\beta_S \label{scs}
\end{equation}
is fulfilled. Moreover, as we pointed out before, it is possible to further extend the thermodynamic space if we interpret the curvature radius $l$ as an independent length-scale in the theory \cite{kubizvnak2017black,dolan2012pdv,kastor2009enthalpy}. Hence, in the most general case, Eq.(\ref{fundame11}) is a quasi-homogeneous function of degree $\beta_M=(D+z-2)/(D-2)\beta_S-(z+1)\beta_l$. Therefore, the cosmological constant is understood as thermodynamic
pressure, $P = -\Lambda/8\pi G$, and a $PV$ term will arise in the first law.



\section{Extended thermodynamics of generalized Lifshitz black holes}
\label{ko}

In this section, we will explore in more detail the extended thermodynamic space of the Lifshitz black hole given by Eqs.(\ref{mko})-(\ref{max}).  In this context, an extended first law is derived and a Smarr relation for the thermodynamic variables. Additionally, for the $3$-dimensional thermodynamic space we will derive the thermodynamic volume for this configuration and check the reverse isoperimetric inequality \cite{cvetivc2011black}.

\subsection{First law in ($S,b)$ equilibrium space}

To have a quasi-homogeneous scaling relation in the thermodynamic system, it is necessary to extend  the equilibrium space. First, we will explore the simplest case, where only $r_h$ and $b$ are independent length scales of the configuration. Let us consider the mass equation
\begin{equation}
    \mathcal{M}=\frac{(D-2)w}{2\kappa l^{z+1}}\left[   1+br_h^{-2(D+z-3)}\right]{r_h}^{D+z-2},
    \label{masa2}
\end{equation}
and taking the variation for two nearby field configurations, we have
\begin{equation}
     d\mathcal{M}=\frac{(D-2)w}{2\kappa l^{z+1}}\Bigg\{\left[ (D+z-2)r_h^{(D+z-3)} -(D+z-4) br_h^{-(D+z-3)}\right] \delta r_h+r_h^{-(D+z-4)}\delta b\Bigg\} \label{mx}.
\end{equation}
From Eq.(\ref{entro}) the variation of the entropy reads
\begin{equation}
    dS=\frac{2 \pi (D-2) w}{k} r_h^{D-3} \delta r_h. \label{entr}
\end{equation}
Next, computing the $TdS$ term using  Eq.(\ref{temp}), and defining 
 \begin{equation}
     B\equiv \frac{(D-2)w}{2\kappa l^{z+1}} r_h^{-(D+z-4)}.\label{Bs}
 \end{equation}
We can compare the above expressions with the mass variation Eq.(\ref{mx}), and we obtain a consistent first law
\begin{equation}
    dM=TdS+Bdb,\label{ju}
\end{equation}
where the quasi-local mass $\mathcal{M}$ is identified with the thermodynamic energy, the Hawking temperature plays the role of thermodynamic temperature, and $B$ is the conjugate variable of $b$, i.e.,
\begin{equation}
    T=\Big(\frac{\partial M}{\partial S}\Big)_b, \quad  B=\Big(\frac{\partial M}{\partial b}\Big)_S.
\end{equation} 
Additionally, from the scaling relation Eq.(\ref{scs}), and using the Euler theorem for quasi-homogeneous functions \cite{belgiorno2003quasi,belgiorno2002notes},  we can pair the first law Eq.(\ref{ju}) with the following Smarr-like relation
\begin{equation}
    (D+z-2)M= (D-2) TS+2(D+z-3)Bb.\label{smar}
\end{equation}
Therefore, to derive a first law consistent with Eulerian scaling, we must interpret the parameter $b$ as a thermodynamic variable. In contrast,  the standard first law $dM=TdS$ was derived in \cite{herrera2021scalarization}, where scaling arguments were ignored (notice that the contribution of the electric term $Q_e$ is absent). This is in accordance with the argument that we pointed out in Sec. \ref{section2}, that one branch of the vector field is needed just to support the structure of the asymptotically Lifshitz spacetime solution. Thus, it seems natural the fact that this vector field does not affect the bulk thermodynamics, but would affect drastically a
potential dual holographic interpretation \cite{tarrio2011black,taylor2008non,taylor2016lifshitz}. Nonetheless, as was mentioned before, if we make the identification $b\sim \Tilde{q}^2$, the configuration behaves like a charged Lifshitz solution, and we could rewrite Eq.(\ref{ju}) as,
    \begin{equation}
    dM=TdS+\Phi_b d\Tilde{q}, \label{jh}
\end{equation}
where
\begin{equation}
\Phi_b\equiv\bigg(\frac{\partial M}{\partial b}\bigg)\bigg(\frac{\partial b}{\partial \Tilde{q}}\bigg).
\end{equation}
Moreover, from the scaling relation Eq.(\ref{scs}), and noticing from Eq.(\ref{xs}) that in a 2-dimensional thermodynamic space $\beta_{\Tilde{q}}=\frac{1}{2}\beta_b$. We might write the Smarr relation as
\begin{equation}
    (D+z-2)M= (D-2) TS + (D+z-3)\Phi_b \Tilde{q},\label{smar1}
\end{equation}
 which is in agreement with the Smarr formula for a charged Lifshitz solution deduced from different arguments in \cite{brenna2015mass}. One might think that this is the correct Smarr relation for Lifshitz spacetimes. Nevertheless, the mass term $M$ scales as $\lambda^{D+z-2}$,
 but as we mentioned before, for $z=1$, we have a relation inconsistent with the $\lambda^{D-3}$
scaling for AdS spacetime \cite{brenna2015mass}. This is tantamount to assuming that
the quantities in the Smarr relation do not depend on the length-scale $l$, as in asymptotically flat spacetimes. No $PV$ term arises as $l$ is not varied. Next, we will address this case.

\subsection{First law in $(S,b,l)$ equilibrium space}
\label{tr}

Now we are interested in obtaining the quantities of thermodynamic
volume $V$ and mass/enthalpy $M$ in the context of extended thermodynamic equilibrium space, in which
the cosmological constant is understood as thermodynamic
pressure, $P = -\Lambda/\kappa$. The variation of the mass Eq.(\ref{masa2}) in this case reads

 \begin{multline}
     d\mathcal{M}=\frac{(D-2)w}{2\kappa l^{z+1}}\Bigg\{\Big[ (D+z-2)r_h^{(D+z-3)} -(D+z-4) br_h^{-(D+z-3)}\Big] \delta r_h\\+r_h^{-(D+z-4)}\delta b-\frac{(z+1)}{l}\Big[ r_h^{(D+z-2)} +br_h^{-(D+z-4)}\Big] \delta l\Bigg\}. \label{mx2}
\end{multline}
Using the Hawking temperature Eq.(\ref{temp}) and the entropy variation Eq.(\ref{entr}), we obtain a $TdS$ term
\begin{equation}
    TdS=\frac{ (D-2) w}{2k l^{z+1}}\Bigg\{\Big[ (D+z-2)r_h^{(D+z-3)} -(D+z-4) br_h^{-(D+z-3)}\Big] \delta r_h\Bigg\}.\label{tgx}
\end{equation}
In addition, if we identify the cosmological constant with a pressure term, from Eq.(\ref{cosmo}) we have

\begin{equation}
    P\equiv\frac{-\Lambda}{\kappa}=\frac{(D+z-3)(D+z-2)}{2\kappa l^2},\quad dP=-\frac{(D+z-3)(D+z-2)}{\kappa l^3}\delta l.\label{press}
\end{equation}
Inserting Eq. \eqref{tgx} and Eq. \eqref{press} in Eq.(\ref{mx2}), we obtain a extended first law with a $PV$-term
\begin{equation}
    dM=TdS+Bdb+VdP \label{ftg},
\end{equation}
where we identified the thermodynamic volume as
\begin{equation}
    V=\frac{ (D-2)(z+1) w}{2(D+z-2)(D+z-3) l^{z-1}}\Big[r_h^{(D+z-2)} + br_h^{-(D+z-4)}\Big],\label{mnj}
\end{equation}
which up to a minus sign in front of the second term, is the same thermodynamic volume deduced from different arguments in \cite{brenna2015mass} for a charged Lifshitz solution. We must remark that this volume is consistent with the notion of thermodynamic volume because it can be obtained directly from the fundamental equation Eq.(\ref{fundame11}) as the conjugate variable to $P$, i.e.,
\begin{equation}
V\equiv\Bigg(\frac{\partial M}{\partial P}\Bigg)_{S,b}=\Bigg(\frac{\partial M}{\partial l}\Bigg)_{S,b}\Bigg(\frac{\partial l}{\partial P}\Bigg)_{S,b}.\label{mnb}
\end{equation}
Notice that for $D\geq 4$, $z>1$, and $b>- r_h^{2(D+z-3)}$ the above volume is always positive. We might compare the thermodynamic volume with the geometric volume, which is essentially defined as 
the full $D$-dimensional volume element over a $t = const$ slice \cite{parikh2006volume,kubizvnak2017black}. For a stationary planar  black hole, it yields
\begin{align}
    V_{geom}&=\int^{r_h}_0 dr\int \sqrt{\abs {g_D}}dx^{D-2}\\ 
    V_{geom}&=\frac{w}{(D-1)}r_h^{D-1}.
\end{align}
It is well known that starting from the charged-AdS black
hole spacetime and employing the definition Eq.(\ref{mnb}), the thermodynamic volume
coincides with the geometric volume \cite{kubizvnak2017black}. Accordingly, it can be noted that in a generalized 
EMD-Lifshitz black hole configuration, the thermodynamic volume Eq.(\ref{mnj}) for $b=0$ and $z=1$ reduces to the geometric volume, i.e., 
\begin{equation}
    V(b\rightarrow0,z\rightarrow1)=\frac{w}{(D-1)}r_h^{D-1}.
\end{equation}
However, in the presence of rotation, additional charges, and other thermodynamic parameters, the formula for the thermodynamic volume gets more complicated, and in general the thermodynamic volume does not seem to have a geometrical meaning. 
In addition, we can relate these results with the first law in the enthalpy representation for an ordinary thermodynamic system
\begin{equation}
    \delta H= T\delta S+V\delta P+\ldots,
\end{equation}
and we can observe that in the presence of a cosmological constant, the mass $M$ has no longer the meaning of internal energy. Rather, $M$ can be interpreted as a gravitational version of chemical enthalpy 
\cite{kastor2009enthalpy}. Furthermore, to match in the isotropic limit $z\rightarrow 1$ the well known $\lambda^{(D-3)}$ scaling relation for AdS spacetime \cite{brenna2015mass}, we impose an extra condition, namely
\begin{equation}
   \beta_l=\frac{ \beta_S}{(D-2)}; \label{gb1}
\end{equation}
and together with Eq.(\ref{scs}), we have that  $M(\lambda^{\beta_S}S,\lambda^{\beta_l}l,\lambda^{\beta_b}b)=\lambda^{(D-3)/(D-2)\beta_S}M(S,l,b)$. Using the above scaling relation and Euler theorem, we have the following Smarr relation for the thermodynamic variables 
\begin{equation}
    (D-3)M= (D-2) TS-2 PV+2(D+z-3) Bb;
\end{equation}
if we make the identification  $b\sim \Tilde{q}^2$, then from Eq.(\ref{xs}) in a 3-dimensional thermodynamic equilibrium space we have $\beta_b=2\beta_{\Tilde{q}}+2z\beta_l$, and using Eq.(\ref{scs}) and Eq.(\ref{gb1}), we have $\beta_{\Tilde{q}}=(D-3)\beta_S/(D-2)$. Therefore, the above relation is nothing more than the Smarr relation for charged  AdS black holes with a $PV$- term \cite{kubizvnak2017black,brenna2015mass}
\begin{equation}
    (D-3)M= (D-2) TS-2 PV+(D-3)\Phi_b \Tilde{q}.\label{smar2}
\end{equation}
Moreover, if we relate the volume Eq.(\ref{mnj}), pressure Eq.(\ref{press}), entropy Eq.(\ref{entro}), and temperature Eq.(\ref{temp}), we obtain the following equation of state
\begin{equation}
    V=\frac{(D-2)(z+1)S}{2P}\Bigg[\frac{r_h^z+b r_h^{-(2D+z-6)}}{4\pi l^{z+1}}\Bigg].
\end{equation}
It is well known that for  asymptotically uncharged planar or
toroidal AdS black holes, the general form of the equation
of state is equivalent to that of an ideal gas $PV = NT$ \cite{altamirano2014thermodynamics}, where $N$ is identified with the
number of degrees of freedom associated with the black
hole, which is proportional to the entropy, hence $N\sim S$. Then, it is straightforward
to check that when $b = 0$ the above relation reduces to the ideal gas law for an uncharged planar Lifshitz solution \cite{brenna2015mass}
\begin{equation}
     V=\frac{(D-2)(z+1)TS}{2(D+z-2)P}.\label{ideal}
\end{equation}
Notice that, as was pointed in \cite{brenna2015mass}, the uncharged\footnote{For $b=0$ the Smarr relation Eq.(\ref{smar}) reduces to $(D+z-2)M= (D-2)TS$ which is the relation first reported in \cite{taylor2008non} and obtained in different Lifshitz spacetime solutions \cite{hyun2015scaling,liu2014thermodynamics,bravo2022lifshitz}.} Smarr relation Eq.(\ref{smar}) is a special case of Eq.(\ref{smar2}),  once the ideal gas law Eq.(\ref{ideal}) is recognized.

\subsection{Reverse isoperimetric inequality}

It is natural to ask if the quantity $V$, defined by Eq.(\ref{mnj}), obeys properties that one would like to associate with the volume of a black hole. A characteristic property of the volume of a simply connected domain
in Euclidean space is that it obeys an isoperimetric inequality. We investigate this
property for the thermodynamic volume in this subsection. In a $(D-1)$-dimensional Euclidean space, the isoperimetric inequality for the volume $V$ of a connected
domain with area  $A$ states that the ratio
\begin{equation}
    \mathcal{R}=\Bigg(\frac{(D-1)V}{w_{D-2}}\Bigg)^{\frac{1}{D-1}}\Bigg(\frac{w_{D-2}}{A}\Bigg)^{\frac{1}{D-2}},\quad \text{where}  \quad w_{D}=\frac{2 \pi^{\frac{D+1}{2}}}{\Gamma(\frac{D+1}{2})},
\end{equation}
is always $\mathcal{R}\leq 1$. Equality holds if and only if the domain is a standard
round ball. It was conjectured in \cite{cvetivc2011black} that a reverse isoperimetric inequality
$\mathcal{R}\geq 1$ holds for any asymptotically AdS black hole, upon identifying $A$ with the horizon area and $V$ with the associated thermodynamic volume, the bound being saturated for
Schwarzschild-AdS black holes. This inequality is essentially
a statement about the amount of entropy a given black
hole can contain. If the ratio is greater than one, the
conjecture implies that the maximal amount of entropy
for that volume has not yet been reached. For the Lifshitz black hole\footnote{For the planar configuration we have to use $w\equiv l^{2-D}\int d^{D-2}x$.} we have
\begin{equation}
    \mathcal{R}=\Bigg[\frac{(D-2)(D-1)(z+1)l^{1-z}}{2(D+z-3)(D+z-2)}\Bigg(r_h^{z-1}+br_h^{-(2D+z-5)}\Bigg)\Bigg]^{\frac{1}{D-1}};
\end{equation}
from the above relation, we can observe that for the $D$-dimensional Schwarzschild  AdS limit, i.e., $z\rightarrow 1$, and $b\rightarrow 0$ we have, as expected, $\mathcal{R}=1$. 
\begin{figure}[H]
\includegraphics[width=0.5\linewidth]{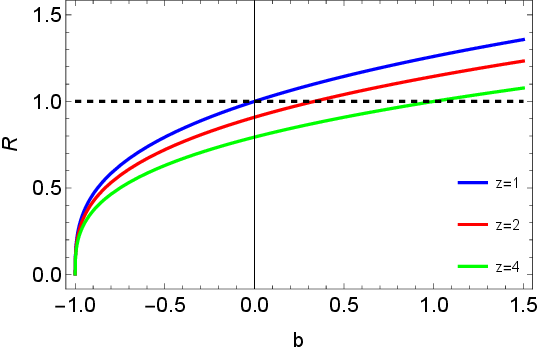}
 \centering
    \caption{$\mathcal{R}$ plotted versus the parameter $b$ for fixed $r_h=1$, $l=1$, $D=4$, and different $z$. There always exists a range for $b>0$ where $\mathcal{R}\geq1$. A similar behavior is observed for $D>4$, $z>1$ and $l>0$.} \label{reverse}
\end{figure}
   
From Fig. \ref{reverse} it can be noted that for $b>0$, there always exists a range where the conjecture is fulfilled. However, for $b< 0$, there is a strong violation of the isoperimetric conjecture, and the configuration behaves as a super-entropic black hole \cite{kubizvnak2017black}. Moreover, for $b<- r_h^{2(D+z-3)}$ the volume becomes negative and $\mathcal{R}$ is not defined. 

 \section{Thermodynamic stability of generalized EMD-Lifshitz black holes}
 \label{sta}
 
Here we explore the thermodynamic stability of the black hole solution, from a local and global perspective. According to classical 
thermodynamics, the local thermodynamic stability, and consequently, the phase transition structure is determined by the behavior of the heat capacity \cite{callen1998thermodynamics}. A more general structure is obtained by considering all non-trivial response functions of the system. Heat capacities are directly related to the heat transfer with the surroundings, and for different processes, one has different heat capacities \cite{avramov2024thermodynamic}. Using the Nambu bracket notation \cite{mansoori2015hessian}, we can provide a general definition of a heat capacity of a thermodynamic system at a fixed set of thermodynamic parameters ($x^1,x^2,\ldots, x^{n-1}$), in a space of variables ($y^1,y^2,\ldots, y^{n}$), namely
\begin{align}
C_{x^1,x^2,\ldots, x^{n-1}}&=T\Bigg(\frac{\partial S}{\partial T}\Bigg)_{x^1,x^2,\ldots, x^{n-1}}=T\frac{\{S,x^1,x^2,\ldots, x^{n-1}\}_{y^1,y^2,\ldots, y^{n}}}{\{T,x^1,x^2,\ldots, x^{n-1}\}_{y^1,y^2,\ldots, y^{n}}}. \label{nambu1}
\end{align}
Notice that the bracket used above generalizes the Poisson bracket for three
or more independent variables. Additionally, 
the relevant state quantities become functions of the coordinates ($y^1, y^2, ..., y^n$) of the space of equilibrium states. The
identification of local stability with the sign of certain heat capacity is related to the
components of the Hessian, where imposing the generic conditions for stability always require
$C > 0$. Heat capacities identify critical and phase structures in the system.
Specifically, if a given heat capacity diverges or changes sign, this would signal the presence of
a phase transition and the breakdown of the equilibrium thermodynamic description; these curves are known as Davies curves or critical curves\footnote{First introduced in the context of black holes in \cite{davies1977thermodynamic}.}. Therefore, if we interpret a black hole as a thermodynamic system, it is natural to infer that the classical condition for local thermodynamic stability is,
\begin{equation}
C_{x^1,x^2,\ldots, x^{n-1}}>0.
\end{equation}
In the forthcoming sections, we are going to revisit the thermodynamic stability of the Lifshitz black hole  in light of the classical criteria presented
above.

\subsection{Local thermodynamic stability in $(S, b)$ space}

\begin{figure}[H]
\begin{minipage}[t]{0.45\linewidth}
 \centering
\hspace{1cm}
\includegraphics[width=1\linewidth]{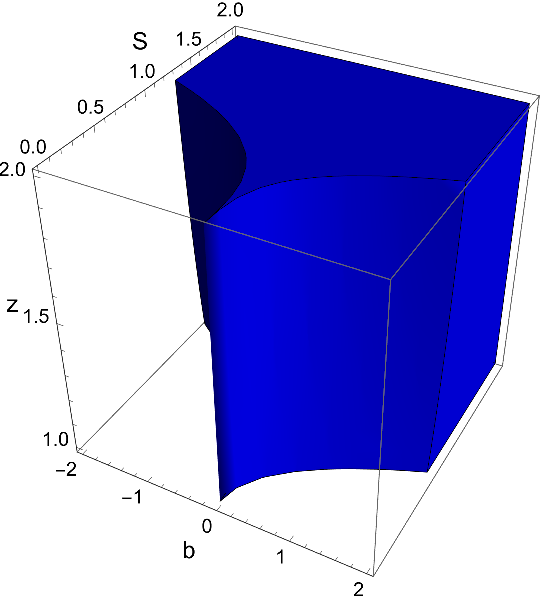}
 (a)\hspace{10cm}
\end{minipage}%
\hfill%
\begin{minipage}[t]{0.5\linewidth}
 \centering
\hspace{1cm}
\includegraphics[width=1\linewidth]{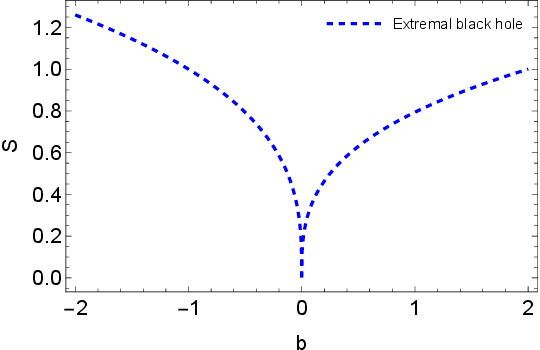}
(b)\hspace{12cm}
\end{minipage}
    \caption{Configuration space for the black hole solution for fixed $w=1$, $k=1$, $l=1$, and $D=4$. (a) The existence of the 4-dimensional black hole is defined inside the blue surface. (b) 2-dimensional view for $z=2$. The solution is defined above the dashed curve. For $b>0$ the curve corresponds to the extremal case, i.e., $T=0$. For $b<0$, the curve corresponds to the non-physical case $M=0$. Similar curves are obtained for different values of $D>3$.}\label{AFM211}

\end{figure}

The local thermodynamic stability of the black hole is determined by the sign of the admissible heat capacities of the solution in the $(S,b)$ space. Using Eq.(\ref{nambu1}) in a two-dimensional equilibrium space, we obtain the following non-trivial heat capacities\footnote{Notice that, $C_S=C_B=0$, and $C_T=\infty$.}
\begin{align}
C_b&=\frac{\Tilde{w}S^{(3D+z-2)/(D-2)}4\pi l^{z+1}T}{a_1S^{2(D+z)/(D-2)}+a_2bS^{6/(D-2)}},
\label{zxv}\\
C_M&=\frac{\Tilde{w}S^{(3D+z-2)/(D-2)}4\pi l^{z+1}T}{a_3S^{2(D+z)/(D-2)}+a_4bS^{6/(D-2)}}.
\end{align}
In the above expressions, we have defined $\Tilde{w}\equiv \frac{2 \pi w}{\kappa}(D-2) $, the entropy variable $\kappa S /2 \pi w\equiv S$, and  the following constants 
\begin{align}
    a_1&=z(D+z-2), \quad a_2=(D+z-4)(2D+z-6),\label{mjks}\\
    a_3&= (D+z-2)(D+2z-4),\quad a_4=(D-2)(D+z-4)\nonumber.
\end{align}
Moreover, in analogy to standard thermodynamics, we can define the compressibility parameters $\kappa_x$ and the coefficient of thermal expansion $\alpha_x$ in the following way
\begin{align}
\kappa_{x}&=-\frac{1}{b}\Bigg(\frac{\partial b}{\partial B}\Bigg)_{x}=-\frac{1}{b}\frac{\{b,x\}_{S,b}}{\{B,x\}_{S,b}},\\
 \alpha_{x}&=\frac{1}{b}\Bigg(\frac{\partial b}{\partial T}\Bigg)_{x}=\frac{1}{b}\frac{\{b,x\}_{S,b}}{\{T,x\}_{S,b}},
\end{align}
which for our fundamental equation 
become\footnote{For this system $\kappa_S=\infty$ and $\alpha_S=\alpha_B$.}
\begin{align}
\kappa_{T}&=\frac{4\pi l^{z+1}S^{(D+z-10)/(D-2)}\bigg[a_1S^{2(D+z)/(D-2)}+a_2bS^{6/(D-2)}\bigg]}{\Tilde{w}b(D+z-4)^2},\\
\alpha_{B}&=-\frac{4 \pi l^{z+1}S^{(2D+z-6)/(D-2)}}{b(D+z-4)}\label{alx}.
\end{align}
It is easy to check that these coefficients are related to the heat capacities as follows:
\begin{equation}
    C_b-C_B=\frac{T b \alpha_B^2}{\kappa_T},
\end{equation}
which is the same well-known relation  for $C_p$ and $C_v$ of standard thermodynamics \cite{callen1998thermodynamics}, where $B$ is the conjugate variable to $b$. In this generalized  EMD-Lifshitz black hole solution, there is no apriori restriction on the new parameter $b$ \cite{herrera2021scalarization}. However, to consider only physical solutions, i.e., configurations with $M>0$ and $T>0$ for arbitrary spacetime dimensions $D$ and free dynamical exponent $z>1$, we have to impose the following constraint (see Fig. \ref{AFM211}) on the thermodynamic variables

 \begin{equation}
  - 1< \frac{b}{S^{2(D+z-3)/(D-2)}}<\frac{(D+z-2)}{(D+z-4)}.\label{restri}
 \end{equation}
 
\begin{figure}[H]
\includegraphics[width=0.6\linewidth]{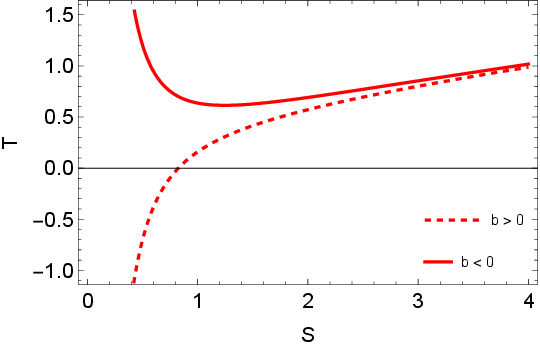}
 \centering
    \caption{Temperature for fixed $w=1$, $k=1$, $l=1$, $z=2$, and $D=4$. For $b>0$, there is a non-physical region where $T<0$. For $b<0$, we have a region where the temperature decreases with the entropy (small unstable black hole) and a region where the temperature increases with the entropy (large stable black hole).} \label{temper1}
\end{figure}

\begin{figure}[H]
\begin{minipage}[t]{0.47\linewidth}
 \centering
\hspace{1cm}
\includegraphics[width=1\linewidth]{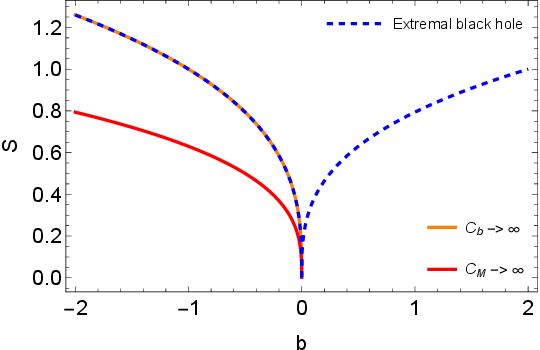}
 (a)\hspace{10cm}
\end{minipage}%
\hfill%
\begin{minipage}[t]{0.47\linewidth}
 \centering
\hspace{1cm}
\includegraphics[width=1\linewidth]{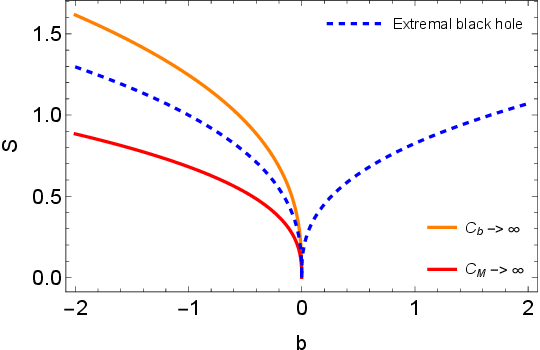}
(b)\hspace{12cm}
\end{minipage}
    \caption{Critical curves for fixed $w=1$, $k=1$, $l=1$, and $z=2$. (a) 4-dimensional case. The black hole is locally stable in the entire physical space. (b) 5-dimensional case. For $b>0$, the black hole is locally stable with respect to both heat capacities in the entire region of existence. For $b<0$, the black hole is only stable above the orange curve.}\label{fgt2}
\end{figure}

The critical curves (or Davies curves) are defined by the divergence of the heat capacity, or when $C_x=0$ (indicating a change of sign in $C_x$), where the configuration changes from unstable to stable or vice versa. For the Lifshitz black hole, both heat capacities are zero, only when $T=0$, which is forbidden by the third law of thermodynamics\footnote{For a review on how it is possible to violate the third law in black holes, see \cite{davies1978thermodynamics,belgiorno2004black}.}. Hence, for $b>0$, both critical curves coincide with the extremal curve (see Fig. \ref{fgt2}), and the black hole is locally stable in this region of the physical space (above the blue curve). Conversely, for $b<0$, $C_b$ and $C_M$ diverge when  $b=\frac{-a_1}{a_2}S^{2(D+z-3)/(D-2)}$ or $b=\frac{-a_3}{a_4}S^{2(D+z-3)/(D-2)}$, respectively\footnote{For an arbitrary value of $D>3$ and $z>1$, the coefficients $a_1, a_2,a_3,$ and $a_4$ are strictly positive}. Consequently, to have a possible phase transition in the system, the coupling parameter $b$ must be negative. This is evident from the behavior of the temperature equation Eq.(\ref{temp}), as illustrated in Fig. \ref{temper1}. 

From Fig. \ref{temper1} we can observe that for $b>0$ the temperature is a monotonic function and,  consequently, no phase transition is expected. Conversely, for $b<0$, there is a local minimum in the temperature that corresponds to the divergence of the heat capacity $C_b$, when the system changes from unstable to stable.
Additionally, observe from Fig. \ref{fgt2} that the critical curve concerning $C_M$ lies in a forbidden region where no black hole solution exists. Therefore, we conclude that throughout the physical region of existence, the black hole is locally stable with respect to processes with constant mass. 

Fig. \ref{fgt2} shows the critical curves for both the $4$-dimensional and $5$-dimensional configurations. In the $4$-dimensional scenario, the critical curve of $C_b$ aligns perfectly with the extremal black hole curve across all values of the dynamical exponent $z$, ensuring local thermodynamic stability for this kind of process. Nonetheless, for $D>4$  we consistently find that the critical curve of $C_b$ lies above the extremal curve for negative couplings, indicating thermodynamic instability in higher-dimensional configurations. At the critical curve, the black hole undergoes a phase transition, breaking down the classical equilibrium description.

\subsection{Global thermodynamic stability in $(S,b)$ space}
According to the general theory, the sufficient conditions for having a stable global equilibrium
in the mass-energy representation is that the Hessian of the energy $\mathcal{H}$ be a positive definite quadratic form. This means that for
global equilibrium, it is sufficient that all eigenvalues of $\mathcal{H}$ be strictly positive \cite{avramov2024thermodynamic}. For this black hole configuration, in a ($S,b$) equilibrium space, the two eigenvalues of the Hessian are
\begin{equation}
    \lambda_{1,2}=S^{-(3D+z-2)/(D-2)}\Bigg[\frac{\kappa U(S)\pm \sqrt{16 \pi^2 w^2 a_4^2 S^{2(D+4)/(D-2)} +\kappa^2U^2(S)}\big]}{16\kappa \pi^2 w(D-2)l^{z+1}}\Bigg],\label{eigen}
\end{equation}
where we have defined the function $U(S)$ as
\begin{align}
U(S)\equiv a_1 S^{2(D+z)/(D-2)}+a_2bS^{6/(D-2)},
\end{align}
for $a_1-a_4$ given by Eq.(\ref{mjks}). Notice that $D>3$ and $z>1$ implies $U(S)>0$, as we analyzed above for the denominator of the heat capacity $C_b$. Thus, for the eigenvalue with a plus sign, we always have $\lambda_1>0$. However, because $c_1$ is a positive constant, then $\lambda_2<0$. Therefore, this indicates that the EMD-Lifshitz black hole cannot be globally stable in a $(S,b)$ space. An alternative set of sufficient conditions for global thermodynamic stability is given by the Sylvester criterion \cite{avramov2024thermodynamic}, which  requires that all of the principal minors of the Hessian of the energy must
be strictly positive, namely
\begin{align}
    \frac{\partial^2 M}{\partial S^2}>0, \quad   \frac{\partial^2 M}{\partial b^2}>0, \quad\abs{\mathcal{H}}>0;
\end{align}
where $\abs{\mathcal{H}}$ is the determinant of $\mathcal{H}$. In the entire physical region of existence Eq.(\ref{restri}), we always have $\partial^2 M/\partial S^2>0$. However, the second condition is never satisfied, because $\partial^2 M/\partial b^2=0$. Moreover, if we compute $\abs{\mathcal{H}}$, we obtain a negative quantity
\begin{align}
 \abs{\mathcal{H}}=-\frac{(D+z-4)^2S^{-2(D+z-6)/(D-2)}}{16\pi^2l^{2(z+1)}}<0.
\end{align}
Therefore, the Sylvester criterion for global thermodynamic stability cannot be fulfilled, and this confirms that the black hole solution is globally unstable.

\subsection{Local thermodynamic stability in $(S,b,l)$ space}
\begin{figure}[H]
\begin{minipage}[t]{0.45\linewidth}
 \centering
\hspace{1cm}
\includegraphics[width=1\linewidth]{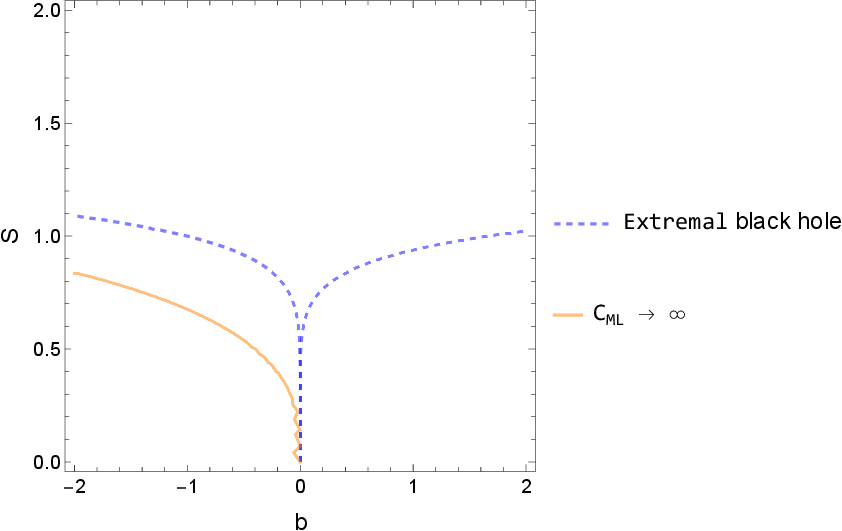}
 (a)\hspace{10cm}
\end{minipage}%
\hfill%
\begin{minipage}[t]{0.5\linewidth}
 \centering
\hspace{1cm}
\includegraphics[width=1\linewidth]{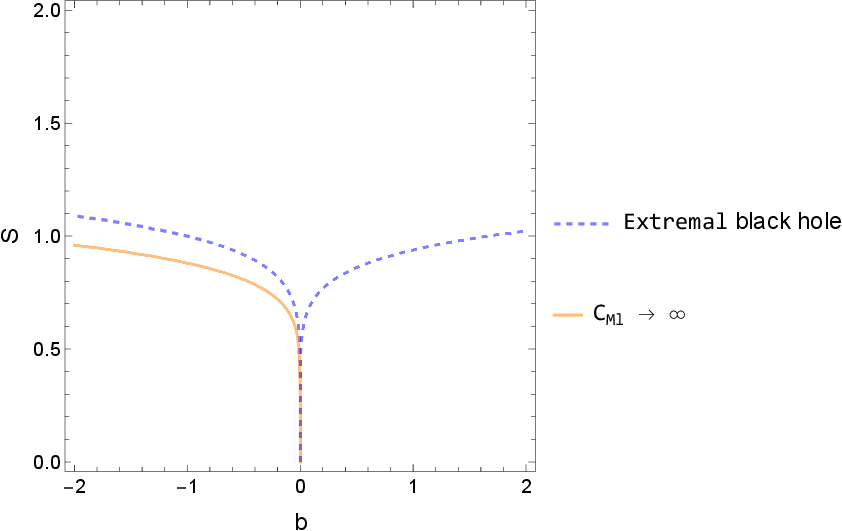}
(b)\hspace{12cm}
\end{minipage}
\hfill%
\begin{minipage}[t]{0.5\linewidth}
 \centering
\hspace{1cm}
\includegraphics[width=1\linewidth]{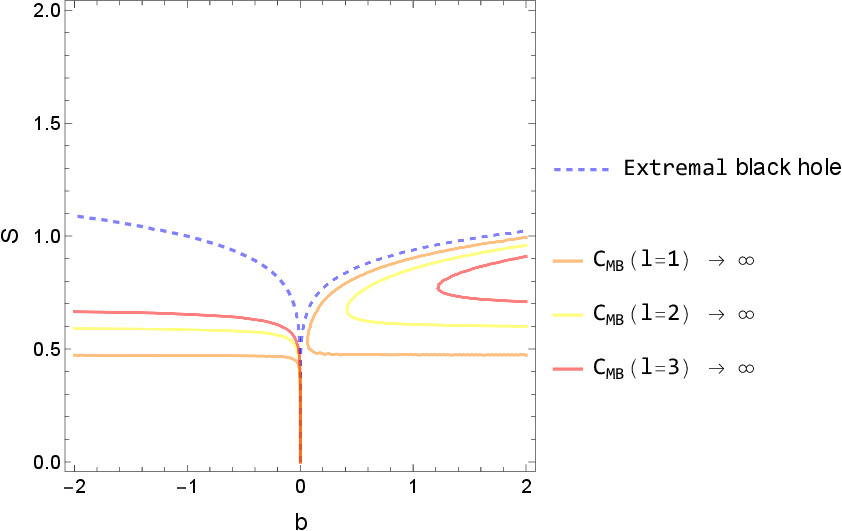}
(c)\hspace{8cm}
\end{minipage}
\hfill%
\begin{minipage}[t]{0.5\linewidth}
 \centering
\hspace{1cm}
\includegraphics[width=1\linewidth]{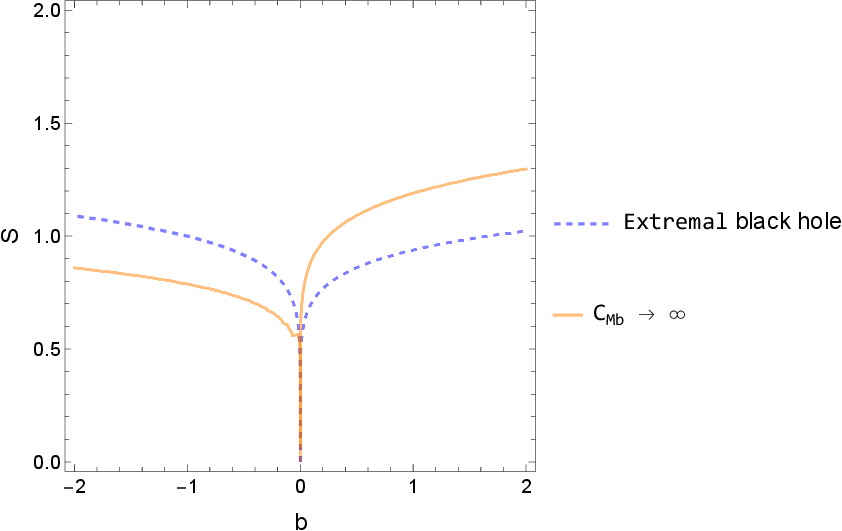}
(d)\hspace{9cm}
\end{minipage}
    \caption{Critical curves of heat capacities at constant mass for fixed $\kappa=1$, $w=1$, $z=2$, and $D=5$. (a) Critical curve of $C_{ML}$. (b) Critical curve of $C_{Ml}$. (c)  Critical curves of $C_{MB}$ for different values of $l$. (d) Critical curve of $C_{Mb}.$}\label{crit1}
\end{figure}

In a $3$-dimensional equilibrium space, there exists a large number of independent heat capacities, due to the large number of state variables involved. For processes at constant mass, we have the following set of heat capacities
\begin{align}
C_{M,L}=\frac{  S F_1 (S,b)}{D_1(S,b)},\quad
C_{M,l}=\frac{S F_1 (S,b) }{ D_2(S,b)},
\end{align}
\begin{align} C_{M,B}=\frac{lS^{D/(D-2)}F_1(S,b)}{D_3(S,b,l)},\quad
C_{M,b}=\frac{ SF_1(S,b)F_2(S,b)}{D_4(S,b)},
\end{align}
where for the sake of simplicity in the above expressions, we have defined 
\begin{align}
F_1(S,b)&=2 \pi  w\Big[a_5S^{2(D+z)/(D-2)}-b a_4 S^{6/(D-2)} \Big],\\
F_2(S,b)&=\Big[S^{2(D+z)/(D-2)}+b S^{6/(D-2)} \Big],\\
D_1(S,b)&= \kappa a_1 S^{2(D+z)/(D-2)}+S^{4/(D-2)}\Big[2\pi w a_4S^{D/(D-2)}+b\kappa a_2 S^{2/(D-2)} \Big],\\
D_2(S,b)&= \kappa a_3 S^{2(D+z)/(D-2)}+b \kappa a_4 S^{6/(D-2)},\\
D_3(S,b,l)&= S^{2(D+z)/(D-2)}\Bigg[2 \pi w a_6 S^{D/(D-2)}+a_7 \kappa l S^{2/(D-2)} \Bigg]\\&+b  S^{6/(D-2)}\Bigg[2 \pi w a_8 S^{D/(D-2)}+\kappa l a_2 S^{2/(D-2)} \Bigg],\nonumber\\
D_4(S,b)&= \kappa \Bigg[ b^2 a_4 S^{12/(D-2)}-a_5 S^{4 (D+z)/(D-2)} +2b a_9 S^{2(D+z-3)/(D-2)}\Bigg],\\
D_5(S,b)&= \kappa\Bigg[a_3S^{2(D+z)/(D-2)}+ba_4 S^{6/(D-2)}\Bigg],\\
D_6(S,b)&=\kappa \Bigg[a_{10}S^{(D+2z+2)/(D-2)}+ba_{11}S^{(8-D)/(D-2)}\Bigg],\\
D_7(S,b)&=\kappa \Bigg[a_1S^{(D+2z+2)/(D-2)}+a_2S^{(8-D)/(D-2)}\Bigg].
\end{align}
The constants $a_1 - a_4$ are given by Eq.(\ref{mjks}) and, additionally, we have defined
\begin{align}
    a_5&= (D-2) (D+z-2), \quad a_6= (D-2)(z+1)(D+z-2),\\
    a_7&=24+2 D(D-7)-16z+5Dz+3z^2, \quad a_8=(z+1)a_4,\\
    a_9&=20+2D^2+2z(z-6)+D(4z-13),\\
    a_{10}&=48+D^2(3z+4)+z[2+z(4z-17)]+D[z(7z-11)-28],\\
    a_{11}&=(D+z-4)[-6-z+D(z+2)].
\end{align}
Notice that in the above expressions only $D_3$ depends explicitly on $l$. Thus, except for $C_{MB}$, the critical curves lie on a 
two-dimensional $(S,b)$ space. This is because, in a Lifshitz black hole with planar horizon\footnote{For a different horizon topology, we expect a non-trivial behavior in $l$ for the heat capacities.},  the curvature radius $l$ appears only as a conformal factor in the fundamental equation Eq.(\ref{fundame11}). Moreover, the physical region of existence of the black hole Eq.(\ref{restri}) is independent of $l$. In Fig. \ref{crit1}, the  $5$-dimensional critical curves for processes at constant mass are plotted.
From Fig. \ref{crit1}, we can observe that for $C_{ML}$, $C_{Ml}$, and $C_{MB}$, the critical curves lie in the forbidden region of the configuration. Therefore, as in the two-dimensional case, the black hole remains locally stable under this kind of process. However, in the case of $C_{Mb}$,  for values of $b>0$, there exists a region where the black hole is unstable (between the orange and blue curves). Furthermore, we have the heat capacities with two constant conjugate parameters\footnote{$ C_{B,l}=C_{x,S}=0$ and $C_{x,T}=\infty$.}
\begin{align}
C_{L,l}&=C_{B,b}=\frac{-S F_1(S,b)}{D_5(S,b)},
\end{align}
and  the set of  heat capacities with constant mixed parameters are
\begin{align}
    C_{L,b}&=-(z+2)l^{-1} S^{-6/(D-2)}F_2(S,b)C_{L,l},\quad
  C_{L,B}=\frac{ (z+1)F_1(S,b)}{D_6(S,b) },\\
C_{b,l}&=\frac{ F_1(S,b)}{D_7(S,b) }.
\end{align}
\begin{figure}[H]
\begin{minipage}[t]{0.45\linewidth}
 \centering
\hspace{1cm}
\includegraphics[width=1\linewidth]{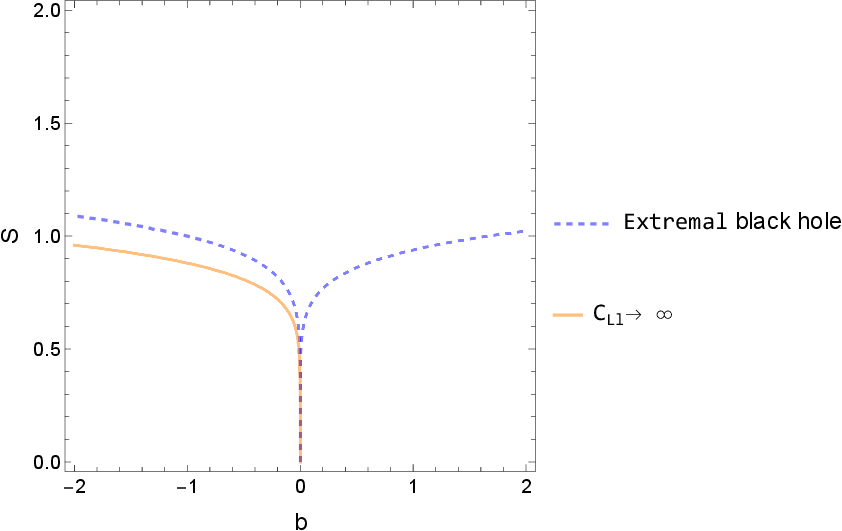}
 (a)\hspace{10cm}
\end{minipage}%
\hfill%
\begin{minipage}[t]{0.5\linewidth}
 \centering
\hspace{1cm}
\includegraphics[width=1\linewidth]{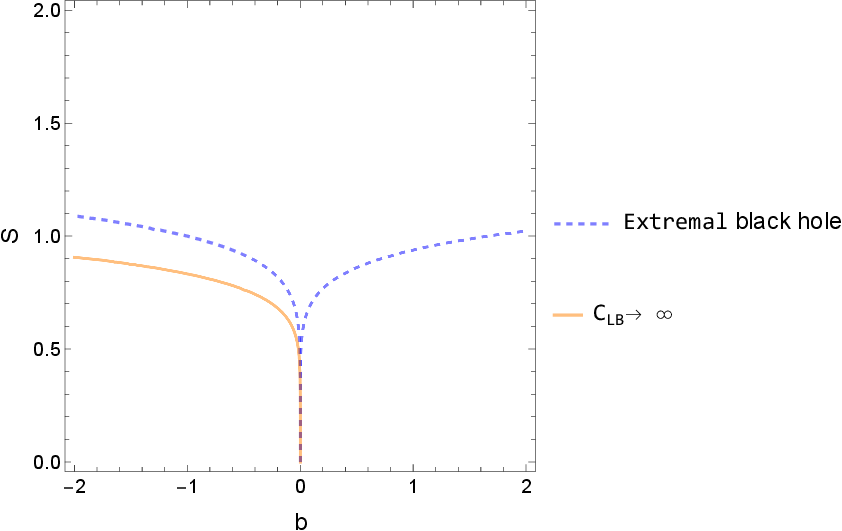}
(b)\hspace{12cm}
\end{minipage}
\hfill%
 \centering
\hspace{1cm}
\includegraphics[width=0.5\linewidth]{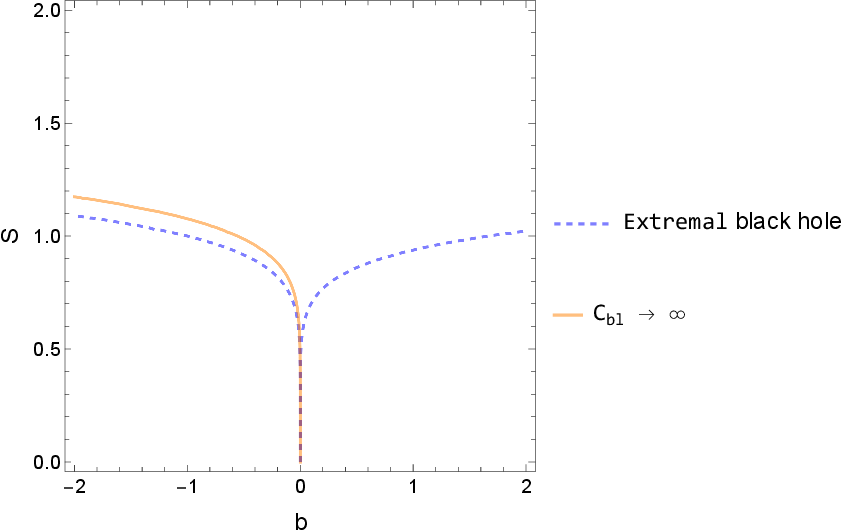}
\hspace{10cm}(c)
    \caption{Critical curves of heat capacities with mixed parameters for fixed $\kappa=1$, $w=1$, $z=2$, and $D=5$. 
 (a) Critical curve of $C_{Ll}$ (b) Critical curve of $C_{LB}$. (c)  Critical curves of $C_{bl}$.}\label{crit2}
\end{figure} 

In Fig. \ref{crit2}, we plot the critical curves for this set of heat capacities. We can observe that only for $C_{bl}$ and values of $b<0$, there exists a region where the black hole is locally unstable.

\subsection{Global  thermodynamic stability in $(S,b,l)$ space}

The global thermodynamic stability is determined by the components of the Hessian
\begin{equation}
    \mathcal{H}(S,b,l)=\begin{pmatrix}
 \mathcal{H}_{SS} & \mathcal{H}_{Sb}  & \mathcal{H}_{lS} \\
\mathcal{H}_{Sb}  & \mathcal{H}_{bb}  & \mathcal{H}_{lb} \\
\mathcal{H}_{lS} &\mathcal{H}_{lb} &\mathcal{H}_{ll} 
\end{pmatrix}=\begin{pmatrix}
M_{,SS}& M_{,Sb}&M_{,Sl} \\
M_{,Sb} & M_{,bb}  & M_{,lb} \\
M_{,Sl}&M_{,lb}&M_{,ll}
\end{pmatrix},
\end{equation}
where $M_{,xy}\equiv \partial^2 M/\partial x \partial y$. The eigenvalues of the Hessian satisfy
a cumbersome cubic equation, which makes it difficult for an analytical treatment. Nevertheless, we can use the conditions imposed by the Sylvester criterion. In the  mass-energy representation, global thermodynamic stability
corresponds to the positive definiteness of the Hessian of the mass. The latter suggests that the first
level principal minors of the Hessian should satisfy
\begin{equation}
    M_{,SS}>0,\quad  M_{,bb}>0,\quad  M_{,ll}>0,\quad 
\end{equation}
together with the conditions for the second-level principal minors:
\begin{equation}
    \Delta_S=\begin{vmatrix}
M_{,bb} & M_{,lb} \\
M_{,lb} & M_{,ll}\end{vmatrix} >0
,\quad   \Delta_b=\begin{vmatrix}
M_{,SS} & M_{,Sl} \\
M_{,Sl} & M_{,ll}\end{vmatrix} >0, \quad \Delta_l=\begin{vmatrix}
M_{,SS} & M_{,Sb} \\
M_{,Sb} & M_{,bb}\end{vmatrix} >0,
\end{equation}
and the determinant of the Hessian itself:
\begin{equation}
    \abs{\mathcal{H}}= \begin{vmatrix}
M_{,SS}& M_{,Sb}&M_{,Sl} \\
M_{,Sb} & M_{,bb}  & M_{,lb} \\
M_{,Sl}&M_{,lb}&M_{,ll}
\end{vmatrix}>0.
\end{equation}
In this case, it only suffices to calculate one minor and show that it is always
negative. Inserting the fundamental equation in the above expressions, we have that
\begin{align}
    \Delta_S&=-\frac{(D-2)^2(z+1)^2w^2S^{-2(D+z-4)/(D-2)}}{4 \kappa^2l^{2(z+2)}}<0,
\end{align}
Hence, as in the $2$-dimensional case, the black hole is globally unstable in a $(S,b,l)$ thermodynamic space. Thus, even the three-parametric thermodynamic
space of equilibrium states for the Lifshitz black hole solution is not enough to support the global thermodynamic stability of the system.

\section{Phase structure and critical behavior of generalized Lifshitz black holes}
\label{sec:crit}

This section aims to construct the phase diagram $b-B$, find critical points, and study the critical behavior of the black hole solution. It is now generally accepted that charged AdS black holes are similar to Van der
Waals fluid systems provided one treats the cosmological
constant as a thermodynamic variable \cite{kubizvnak2017black}. Also, it has been shown in \cite{dehyadegari2017critical}
that the connection between charged AdS black holes and  Van der Waals fluid systems can be made without extending
the phase space \cite{dehyadegari2017critical} by keeping the cosmological
constant as a fixed parameter. The key assumption in this picture is to treat the square of the charge of the black
hole, $Q^2$, as a thermodynamic variable instead of the charge $Q$ \cite{dehyadegari2017critical}. Notice that Lifshitz black holes are anisotropic generalizations of AdS black holes. So, it is natural to expect that the phase structure and critical behavior of Lifshitz black holes be similar to Van der Waals systems as well. 

From the expression of temperature Eq.(\ref{temp}), we can see that for an
arbitrary Lifshitz exponent $z$, and $b\neq 0$ it is impossible to solve for $r_h$ and write an analytical equation of state for $P(V,T)$. This implies that, for Lifshitz dilatonic black holes, one cannot investigate analytically the
critical behavior of the system through an extended $P-V$ equilibrium space by treating the cosmological constant as a thermodynamic variable (pressure). Nevertheless, recently  in \cite{dayyani2018critical}, it was shown that for dilatonic Lifshitz black holes in the presence of a  Maxwell power-law electromagnetic field, it is quite possible to investigate the critical behavior of this system through a new\footnote{The equation of state is given by $ Q^s= Q^s(T,\Psi)$, where $\Psi$ is the
conjugate of $Q^s$,
$s=2p/(2p-1)$, and $p$ is the power of the Maxwell term in the Lagrangian \cite{dayyani2018critical}.} $Q^s-\Psi$ equilibrium space, which reveals the similarity with a Van der Waals fluid system in an ensemble, where the cosmological constant is treated as a fixed
parameter and the charges vary. As was pointed out in Sec. \ref{section2}, the planar black hole solution given by Eqs.(\ref{mko})-(\ref{max}) behaves like a charged Lifshitz black hole because of the presence of the generalized dilatonic coupling parameter $b$. Thus, we will follow a similar approach to the one presented in \cite{dayyani2018critical}, and analyze the critical behavior in a $b-B$ equilibrium space.

\subsection{Gibbs free energy}
\label{Gen}
\begin{figure}[H]
\begin{minipage}[t]{0.48\linewidth}
 \centering
\hspace{1cm}
\includegraphics[width=1\linewidth]{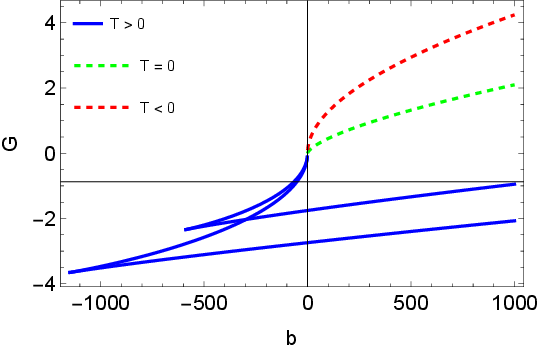}
 (a)\hspace{10cm}
\end{minipage}%
\hfill%
\begin{minipage}[t]{0.48\linewidth}
 \centering
\hspace{1cm}
\includegraphics[width=1\linewidth]{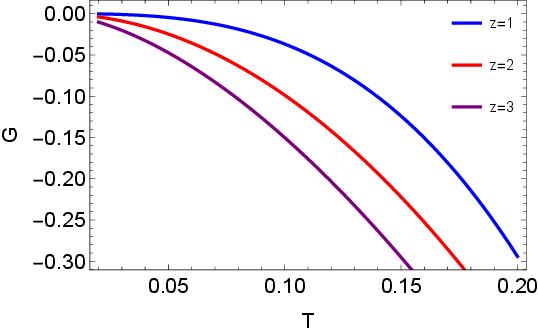}
(b)\hspace{12cm}
\end{minipage}
    \caption{Isotherms of the Gibbs free energy for fixed $z=2$, $D=4$, $l=1$, $w=1$, and $\kappa=1$. (a) Charged case. The system undergoes a first-order phase transition. (b) Uncharged case. The system behaves like an ideal gas. A similar behavior is observed for different values of $D$, $l$, and $z$.}\label{g1}
\end{figure}

The Gibbs free energy $G$ is one of the most important tools that we have to obtain information about the phase transition structure of a thermodynamic system. It is well-known that discontinuities in the $n$-derivative of $G$ indicate $n$-order phase transitions according to the Ehrenfest scheme \cite{callen1998thermodynamics}. For example, a discontinuity in the first derivative of $G$ corresponds to a first-order phase transition, and so on. By regarding the mass as the gravitational enthalpy \cite{kastor2009enthalpy}, the Gibbs free energy is defined as
\begin{equation}
    G(T,l,b)\equiv M-TS=\frac{wl^{-(1+z)}\Big[-zr_h^{(D+z-2)}+br_h^{-(D+z-4)}(2D+z-6)\Big]}{2 \kappa},
\end{equation}
where from Eq.(\ref{temp}) it is impossible to obtain $r_h(T,l,b)$ for a arbitrary values of $D$ and $z$. However, we can plot isotherms for $G$ for different values of $D$, $z$, and fixed $l$.  In Fig. \ref{g1}, we can observe that for the charged case ($b\neq 0$) the typical swallowtail behavior of the free energy
indicates that a first-order phase transition occurs in the system for negative couplings. Conversely, for the uncharged case ($b=0$), we don't have a phase transition structure, in agreement with the ideal gas law Eq.(\ref{ideal}).
\subsection{Equation of state in $b-B$ equilibrium space}
To study the critical behavior of the black hole solution, first we solve from Eq.(\ref{temp}) for the coupling $b$, and treating  $l$ as a fixed parameter. Then, from Eq.(\ref{Bs}),  we can relate $r_h$ and $B$ to obtain the following equation of state 
\begin{equation}
    b(B,T)=\frac{(D+z-2)}{(D+z-4)}\Bigg[\frac{(D-2)w}{2\kappa B l^{z+1}}\Bigg]^{2(D+z-3)\big/(D+z-4)}-\frac{4 \pi l^{z+1}T}{(D+z-4)}\Bigg[\frac{(D-2)w}{2\kappa B l^{z+1}}\Bigg]^{(2D+z-6)\big/(D+z-4)}.\label{critical1}
\end{equation}
To explore critical behavior, we need to obtain the inflection points of the equation of state, namely 
\begin{equation}
    \frac{\partial b}{\partial B}\biggm|_{T_c}=0,\quad \frac{\partial^2 b}{\partial B^2}\biggm|_{T_c}=0;\label{stationpoints}
\end{equation}
which allows the only mathematical solution 
\begin{equation}
    B_c=\infty,\quad T_c=0,\quad b_c=0.
\end{equation}
 Therefore, as expected for planar black holes, we conclude that this black hole solution doesn't exhibit critical behavior, in agreement with \cite{dayyani2018critical} for the planar case and $p=1$, once we identify  $b\sim l^{2z} Q^2$ and $B\sim \psi/l^{z+1}$.

\subsection{Critical behavior of Lifshitz black holes with spherical and hyperbolic horizon}
Recently, in \cite{wu2024thermodynamics}, the 4-dimensional solution for a generalized EMD-Lifshitz black hole with different horizon topologies was obtained. Using these results, we will explore the critical behavior of the  4-dimensional black hole configuration with spherical and hyperbolic horizons. For this case, the mass equation and Hawking temperature are given by
   \begin{align}
    \mathcal{M}=\frac{w}{\kappa l^{z+1}}\left[   {r_h}^{z+2}+\frac{b}{r_h^z}+\frac{\eta l^2 r_h^z}{z^2}\right],
    \label{masa3}\quad
    T=\frac{r_h^{3z+2}\Big[\eta l^2+z(z+2)r_h^2\Big]-b z^2 r_h^{z+2}}{4\pi z l^{z+1}},
 \end{align}
where $\eta$ is the parameter related to the horizon geometry. It is trivial to check that for the 4-dimensional planar case, i.e., $\eta=0$, the above expressions reduce to Eq.(\ref{masa2}) and  Eq.(\ref{temp}), respectively. Notice that  due to the non-trivial horizon topology, the equation of state adds an extra term and reads
 \begin{equation}
    b(B,T)=\frac{z+2}z\Bigg[\frac{w}{\kappa B l^{z+1}}\Bigg]^{2(z+1)/z}+\frac{\eta l^2}{z^2}\Bigg[\frac{w}{\kappa B l^{z+1}}\Bigg]^2-\frac{4 \pi l^{z+1}T}{z}\Bigg[\frac{w}{\kappa B l^{z+1}}\Bigg]^{z+2/z}.\label{critical2}
\end{equation}
From Eq.\eqref{masa3}, it can be observed that $B$ is independent of the horizon geometry and, consequently, it is the same as in the planar case, as given by Eq.(\ref{Bs}). Furthermore, using Eqs.(\ref{stationpoints}), the critical points read
\begin{align}
\begin{aligned}
   B_c&= \frac{w}{\kappa l^{2z+1}}\Bigg[\frac{\sqrt{z (z+1)(z+2)}}{\sqrt{\eta(2-z)}}\Bigg]^z,\quad T_c=\frac{(z+1)}{(2-z)\pi l}\Bigg[\frac{\eta(2-z)}{z (z+1)(z+2)}\Bigg]^{z/2},\\
   b_c&= \frac{l^{2(z+1)}}{z}\Bigg[\frac{\eta(2-z)}{z (z+1)(z+2)}\Bigg]^{z+1}. \label{critical_nhorizon}
   \end{aligned}
\end{align}
\begin{figure}[H]
    \centering
    \includegraphics{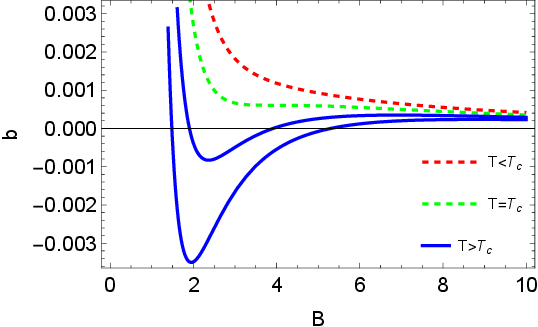}
    \caption{Isotherms $b-B$ for the black hole with spherical horizon for $z=1.4$, $D=4$, $l=1$, $w=4 \pi$, and $\kappa=8\pi$.}
    \label{isoter}
\end{figure}
From Fig. \ref{isoter}, we can observe that for $T>T_c$, the isotherms have an oscillating part, which is typical for a first-order phase transition between a "liquid" and "gas" phase in a van der Waals fluid system, which is supported by the analysis of the free energy performed in Sec. \ref{Gen}. Moreover, when $T=T_c$, this local minimum in the equation of state becomes an inflection point, usually interpreted as a second-order phase transition.

\begin{figure}[H]
\begin{minipage}[t]{0.48\linewidth}
 \centering
\hspace{1cm}
\includegraphics[width=1\linewidth]{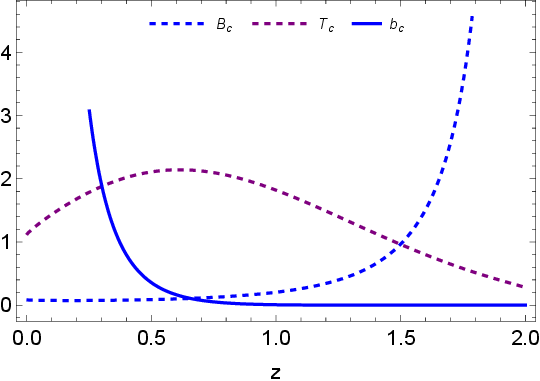}
 (a)\hspace{10cm}
\end{minipage}%
\hfill%
\begin{minipage}[t]{0.48\linewidth}
 \centering
\hspace{1cm}
\includegraphics[width=1\linewidth]{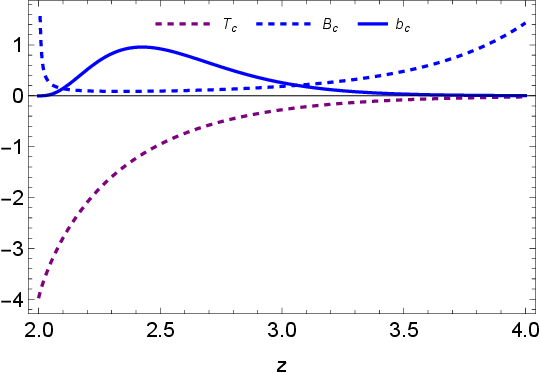}
(b)\hspace{12cm}
\end{minipage}
    \caption{Critical variables for $w=4 \pi$, $k=8 \pi$, $l=1$, and different values of $z$. (a) Spherical horizon. For $0<z<2$, all critical variables are real and positive. However, for $z>2$ they become imaginary.  (b) Hyperbolic horizon. For $z>2$, $T_c$ is negative and for $0<z<2$, all critical parameters are imaginary.}\label{critic_temp}
\end{figure}

Additionally, notice that for the hyperbolic case, i.e., $\eta=-1$, no critical behavior occurs since for values of $z>2$, the critical temperature is negative, as shown in Fig. \ref{critic_temp}. Moreover, for $0<z<2$, all critical variables are imaginary.

Conversely, for a black hole with spherical horizon geometry and dynamical exponent values $0<z<2$, the critical temperature is positive, as depicted in Fig. \ref{critic_temp}. Therefore, we conclude that at least for the 4-dimensional configuration, critical behavior is observed only in the spherical case for values\footnote{The lower bound of $z=1$ is necessary to ensure the correct asymptotic behavior of the metric (\ref{metrica1}).} of $1\leq z<2$ . This is not surprising, since Lifshitz black holes are continuously connected with AdS black holes, and it is well-know that only spherical charged AdS black holes exhibit critical behavior \cite{kubizvnak2017black,kubizvnak2012p}. Moreover, from Eq.(\ref{critical_nhorizon}) we notice that critical behavior is only defined strictly for $z<2$, as for $z=2$ the values of $B_c$ and $T_c$ diverge. This upper bound on the dynamical exponent $z$ has been reported in various studies of Lifshitz black holes using different equations of state. For instance, in EMD-Lifshitz black holes with a hyperscaling violating parameter, using an equation of state given by the inverse temperature as a function of the horizon radius and the electric charge \cite{pedraza2019hyperscaling}, and in an Einstein-Bumblebee-Dilaton framework, using a $P_{LV}-V$ equilibrium space, where $P_{LV}$ is the pressure associated to the Bumblebee field  \cite{lessa2024einstein}. In both configurations, it has been shown that critical behavior is only present for values in the interval $1\leq z<2$, which is in agreement with our results. Thus, it seems that these bounds in the dynamical exponent $z$ are quite general, since they are independent of the equilibrium space and equation of state used to study the thermodynamic properties of the system. In addition, we can use the critical quantities to define a $l$ independent constant, which for the spherical case leads to  
\begin{equation}
    \rho_c\equiv b_c T_c B_c=\frac{w}{\pi \kappa z^2(z+2)}\Bigg[\frac{z(z+1)(z+2)}{2-z}\Bigg]^{-3z/2}\Bigg[\sqrt{\frac{z(z+1)(z+2)}{2-z}}\Bigg]^{z}. \label{energydensi_critical}
\end{equation}
\begin{figure}[H]
    \centering
    \includegraphics{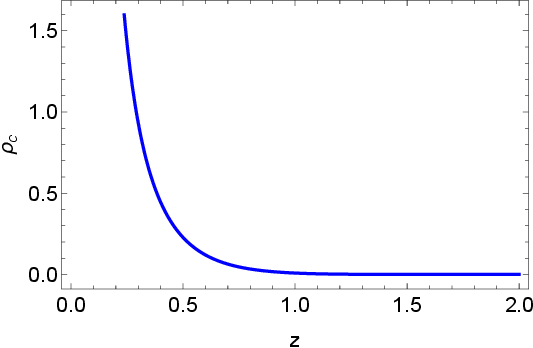}
    \caption{$\rho_c$ diagram for $w=4 \pi$, $k=8 \pi$, and different values of dynamical exponent $z$. $\rho_c$ is real and positive for values of $0<z<2$. For $z>2$ $\rho_c$ is imaginary.}
    \label{critical_energy}
\end{figure}
This suggests that the physical behavior of Lifshitz black holes near the critical point only depends on the value of the dynamical exponent $z$, as illustrated in Fig. \ref{critical_energy}. Notably, for the spherical case and in the isotropic limit $z=1$. The critical points reduce\footnote{We used explicitly the value of $w=4 \pi$ and $\kappa=8 \pi G$, where $G$ is the 4-dimensional Newton's constant.} to

 \begin{align}
     B_c=\sqrt{\frac{3}{2}}\frac{1}{l^3 G}, \quad T_c=\sqrt{\frac{2}{3}}\frac{1}{\pi l}, \quad b_c=\frac{l^4}{36}, \quad \rho_c=\frac{1}{36 \pi G},
 \end{align}
which coincides with the critical points in a $Q^2-\Psi$ equilibrium space for the spherical charged AdS black holes \cite{dehyadegari2017critical}, once we identify $b\sim l^2 Q^2$ and $B\sim \Psi/l^2$.

\subsection{Critical exponents}

The critical exponents are codified in the behavior of the thermodynamic variables in the vicinity of critical points. From the results obtained above, we conclude that Lifshitz black holes with planar and hyperbolic horizon does not present criticality. Nevertheless, for the spherical horizon geometry we expect critical behavior. Hence, in this section we compute the critical exponents \cite{callen1998thermodynamics} $\alpha$, $\beta$, $\gamma$, and $\delta$, using the equation of state in a $b-B$ equilibrium space.  First, we define the reduced variables 
    \begin{equation}
      T_r= \frac{T}{T_c} , \quad B_r= \frac{B}{B_c}, \quad b_r= \frac{b}{b_c}.
\end{equation}
Since by definition we are near the critical point, we write the reduced variables as
\begin{equation}
    T_r=1+t ,\quad B_r=1+\phi,\label{reduce}
\end{equation}
where $\phi$ is the order parameter and $t$ and $\phi$  indicate the  deviation from criticality. Next, we perform a Taylor expansion of the equation of state around the critical point, namely
\begin{multline}
    b=b_c+\frac{\partial b}{\partial T}\biggm|_c(T-T_c)+\frac{\partial b}{\partial B}\biggm|_c(B-B_c)+\frac{1}{2}\Bigg[\frac{\partial^2 b}{\partial T^2}\biggm|_c{(T-T_c)}^2+\frac{\partial^2 b}{\partial B^2}\biggm|_c{(B-B_c)}^2\\+2\frac{\partial^2 b}{\partial T\partial B}\biggm|_c(T-T_c)(B-B_c)+\frac{1}{3}\frac{\partial^3 b}{\partial B^3}\biggm|_c{(B-B_c)}^3+\ldots \Bigg].
\end{multline}
However, $b(B,T)$ is lineal in $T$, i.e, $\partial^2 b/\partial T^2=0 $ and, by the definition of a critical point, we have that  $\partial b/\partial B\biggm|_c= \partial^2 b/\partial B^2\biggm|_c=0$. Therefore, using the reduced variables Eq.(\ref{reduce}), the above expression simplifies to
\begin{equation}
    b_r=1+w_1t+w_2 \phi t+ w_3 \phi^3 + O(t\phi^2,\phi^4); \label{criticalex}
\end{equation}
where we have defined the following constants
\begin{align}
\begin{aligned}
w_1\equiv \frac{\partial b}{\partial T}\biggm|_c \Bigg(\frac{T_c}{b_c}\Bigg)&=\frac{-4(z+1)}{(2-z)},\quad w_2\equiv \frac{\partial^2 b}{\partial T\partial B}\biggm|_c\Bigg(\frac{T_c B_c}{b_c}\Bigg)=\frac{4(z+1)(z+2)}{z(2-z)},\\
w_3\equiv \frac{1}{6}\frac{\partial^3 b}{\partial B^3}\biggm|_c\Bigg(\frac{B_c^3}{b_c}\Bigg)&=\frac{-2(z+1)(z+2)}{3z^2}.\label{coef_criti}
\end{aligned}
\end{align}
For $z=1$, the above coefficients  simplify to those reported in \cite{dehyadegari2017critical}. Notice that the coefficients $w$ are independent of the horizon geometry and, therefore, coincide with those of the spherical and hyperbolic cases, in agreement with the results of \cite{pedraza2019hyperscaling,dayyani2018critical}. To calculate the critical exponent $\alpha$, we consider the behavior of the specific heat at constant $B$ near the critical point \cite{callen1998thermodynamics}
\begin{equation}
    C_B=T\Bigg(\frac{\partial S}{\partial T}\Bigg)_B\propto \abs t^{-\alpha}.
\end{equation}
Using Eq.(\ref{entro}), Eq.(\ref{Bs}), and for $D=4$, we have that the entropy in a $b-B$ equilibrium space is only a function of the variable $B$ and reads
\begin{equation}
    S(B,T)=\frac{2 \pi w}{\kappa}\Bigg[{\frac{ w}{\kappa l^{z+1}B}\Bigg]}^{2/z}.
\end{equation}
Then, the heat capacity $C_B=0$ for all values of temperature. Hence, the critical exponent is $\alpha=0$. Next, to compute the critical exponent $\beta$, we need to analyze the order parameter $\phi$ near criticality. In doing that, we use  Maxwell's equal area law \cite{callen1998thermodynamics} denoting $\phi_l$ and $\phi_g$ as the order parameters for liquid and gas phase, respectively. Then, 
 \begin{align}
 b_r&=1+w_1t+w_2 \phi_lt+ w_3 \phi_l^3=1+w_1t+w_2 \phi_s t+ w_3 \phi_s^3,\nonumber\\
      0&=\int^{\phi_g}_{\phi_l}\phi db_r.\label{max1}
 \end{align}
    Differentiating $b_r$ of Eq.(\ref{criticalex}) with respect to $\phi$ and inserting the result in Eq.(\ref{max1}), we obtain
    \begin{equation}
        (\phi_g-\phi_l)^2\Bigg[\frac{w_2t}{2}+\frac{3}{4}w_3 (\phi_g-\phi_l)^2\Bigg]=0.
    \end{equation}
    The only non-trivial solution for the above equation is
    \begin{equation}
           \phi_l=-\phi_g=\sqrt{\frac{-w_2}{6w_3}}t^{1/2}.
    \end{equation}
 In ordinary thermodynamics, we know that the exponent $\beta$ describes the behavior of $\abs {\phi_l-\phi_s}\propto \abs t ^\beta$ . Consequently, for the Lifshitz black hole, we have
 \begin{equation}
     \abs {\phi_l-\phi_g}=2 \phi_g=\sqrt{\frac{-2w_2}{3w_3}}t^{1/2}.
 \end{equation}
 From the above result, we can read the critical exponent $\beta=1/2$. To calculate the exponent $\gamma$, one might examine the isothermal "susceptibility" near the critical point
 \begin{equation}
     \chi_T=\Bigg(\frac{\partial B}{\partial b}\Bigg)_T\propto \abs t ^{-\gamma}.
 \end{equation}
Differentiating Eq.(\ref{criticalex}) with respect to $\phi$, we obtain
\begin{equation}
    \chi_T=\frac{1}{w_2 t},\quad \implies \gamma=1.
\end{equation}
Finally, the critical exponent $\delta$ governs the behavior of $\abs{b-b_c}\sim \abs{B-B_c}^\delta$. This is equivalent to the expression $\abs{b_r-1}\sim \abs{\phi}^\delta$. Therefore, evaluating Eq.(\ref{criticalex}) in the critical isotherm ($t=0$), we obtain
\begin{equation}
  \abs { b_r-1}\sim \abs{w_3} \abs\phi^3 \quad\implies \delta=3.
\end{equation}
In conclusion, we can observe that using the $b-B$ equilibrium space, the EMD-Lifshitz black hole with a generalized coupling function resembles a van der Waals fluid. Moreover, the critical behavior is described by the mean field theory  critical exponents \cite{landau2013statistical,callen1998thermodynamics}.

\section{Closing remarks}
\label{sec:con}

This work aimed to investigate the properties of a black hole solution of Einstein-Maxwell gravity with cosmological constant and a dilaton field non-minimally coupled to the electromagnetic field, whose spacetime is described by a Lifshitz anisotropic metric.
We assume that the black hole can be considered as a thermodynamic system and use the formalism of quasi-homogeneous and extended thermodynamics to investigate the main physical characteristics. 

First, to derive a first law consistent with scaling arguments, we need to interpret the coupling parameter $b$ as a new thermodynamic variable. Therefore, the fundamental equation of the Lifshitz black hole is described by a quasi-homogeneous function. Moreover, to match in the $z=1$ limit the scaling of the fundamental equation of the AdS black hole, we need to interpret the curvature radius $l$ as a thermodynamic variable, and, consequently, a $PV$ term appears in the first law. Furthermore, we analyze the thermodynamic stability of the solution from the local and global perspective. The configuration is globally unstable in a two- or three-dimensional equilibrium space. However, for the 4-dimensional configuration, the solution is locally stable across the entire physical space. In higher dimensional configurations, the solution retains local stability for processes at constant mass and constant positive couplings, but a region of local instability emerges for negative couplings. In addition, we analyze the phase structure of the black hole solution, and we conclude that in the standard dilatonic setup ($b=0$), the planar black hole behaves like an ideal gas. Conversely, for the generalized case ($b\neq0$), we study the Gibbs free energy in a fixed $l$ ensemble and demonstrate that for negative couplings, the configuration undergoes a first-order phase transition, similar to the liquid/gas phase transition in a van der Waals fluid, which in the context of black holes thermodynamics is known as small/large black hole phase transition \cite{kubizvnak2017black}. Next, we study criticality using an analytical equation of state $b(B,T)$ and observe that planar dilatonic Lifshitz black holes do not show a critical behavior in any number of spacetime dimensions. Thus, generalizing the dilatonic coupling is not sufficient to achieve critical behavior in Lifshitz planar black holes. Nonetheless, the 4-dimensional black hole with spherical horizon exhibit critical behavior for values of $1\leq z<2$. Finally, we computed the critical exponents of the system and found that they are independent of the dynamical exponent $z$ and dilaton coupling function because they are universal and exactly the same as those of the mean field theory \cite{landau2013statistical,callen1998thermodynamics}. Notably, in the isotropic limit we recover the critical points of the spherical charged AdS black hole in a $Q^2-\Psi$ equilibrium space \cite{dehyadegari2017critical}. This means that spherical dilatonic Lifshitz black holes resemble the critical behavior of a van der Waals fluid, without needing to introduce the notion of pressure and thermodynamic volume.

This work can be expanded and applied in various directions. Given that Lifshitz black holes represent anisotropic versions of AdS black holes, we can use the geometrothermodynamics approach \cite{quevedo2007geometrothermodynamics} to explore how the dynamical exponent
$z$ influences the microstructure analysis of charged black holes \cite{ladino2024phase}. Also, we may explore the extended thermodynamics of other non-relativistic backgrounds \cite{taylor2008non,taylor2016lifshitz} such as the Schr\"odinger spacetime within the EMD framework \cite{herrera2021hyperscaling,herrera2023anisotropic}. Moreover, we believe that a more detailed study is needed to fully understand the role of the dilatonic coupling in dual condensed matter systems.
\section*{Acknowledgments}

This work was partially supported by Conahcyt-Mexico, Grant No. A1-S-31269. 
CRF acknowledge support from Conahcyt-Mexico through a PhD grant.

\bibliographystyle{unsrt}
\bibliography{referencias.bib}

\end{document}